# INNOVATIVE RATED-RESOURCE PEER-TO-PEER NETWORK

Abhishek Sharma and Hao Shi

School of Engineering and Science, Victoria University, Melbourne, Australia
`abhisharayiya@gmail.com` and `hao.shi@vu.edu.au`

## ABSTRACT

*Peer-to-Peer (P2P) networks provide a significant solution for file sharing among peers connected to Internet. It is fast and completely decentralised system with robustness. But due to absence of a server documents on a P2P network are not rated which makes it difficult for a peer to obtain precise information in result of a query. In past, some researchers tried to attach ratings to the peers itself but it was complex and less effective. In this paper, a novel P2P architecture is proposed which attaches ratings to the uploaded document directly. These ratings then become as <Rating> element in its XML advertisement which has several child elements for information classification. The attached <Rating> element is extracted from the advertisement in real time and the document is then sorted accordingly. Therefore, the information can be easily sorted based on a request by a peer according to the relevance of matter. The information regarding relevance is obtained by the peer issuing the query. This research leads to a smart P2P model, the Rated-Resource P2P network (R2P2P).*

## KEYWORDS

*P2P, rated-resource, XML, JXTA, R2P2P, LDAP, vuCRN*

## 1. INTRODUCTION

Peer-to-Peer (P2P) technology enables any network-connected device to provide services to another network-connected device. A device in a P2P network can provide access to any type of resource that it has at its disposal, whether documents, storage capacity, computing power, or even its own human operator. The device in a P2P network could be anything ranging from a super computer to simple PDA. P2P technology is a robust and impressive extension of the Internet's philosophy of robustness through decentralization.

The main advantage of P2P networks is that it distributes the responsibility of providing services among all peers on the network; this eliminates service outages due to a single point of failure and provides a more scalable solution for offering services. In addition, P2P networks exploit available bandwidth across the entire network by using a variety of communication channels and by filling bandwidth up to the brim of the Internet. Unlike traditional client/server communications, in which specific routes to popular destinations can become overloaded (for example, the route to google.com), P2P enables communication via a variety of network routes, thereby reducing network overloading. P2P has the capability of serving resources with high availability at a much lower cost while maximizing the use of resources from every peer connected to the P2P network. Where as, client/server solutions rely on the addition of costly bandwidth, equipment, and co-location facilities to maintain a robust solution. P2P can offer a similar level of robustness by spreading network and resource demands across the P2P network. Several different P2P architectures have been proposed so far, a comprehensive survey is provided in [1].

Although, P2P networks provide a wonderful solution to a completely decentralized network there are still some issues that need some attention. One of the major drawbacks in P2P networks is the unsupervised uploading of content. This leaves a room for infecting the network with viruses and worms. This problem was addressed by a research group in Victoria University [2] by supervising and controlling the upload on the network. The problem of piracy and copyright protection was addressed by attaching digital metadata to the uploaded materials [5]. These two implementations tackled the problem of network-infection and Digital Right Management (DRM) [3]. But still there is plenty of room for further improvement in making P2P networks even more useful.

A collaborative research network needs to be precise in providing information sought by a user. So, a system that provides the response to a search query should be intelligent enough to sort them in the order of relevance. This article addresses this issue and provides a possible model of such a system by attaching rating to the documents uploaded on the network. The ratings attached will be issued by the author and could be revised by other members in the network. While planning such a system care is taken to restrict the right of modification in the rating of the document to few people only. The resulting smart model is termed as Rated-Resource P2P network R2P2P.

Section 2 discusses related work. Section 3 and 4 put some light on JXTA and *vuCRN* network. In Section 5 the details of the proposed network architecture is discussed. Finally Section 6 is the conclusion and future work.

## 2. RELATED WORK

There has been a lot of development in this field to make the network a smart one to facilitate better information retrieval for a search query. The approaches can be broadly divided into main categories – Node based and Resource based. While the former refers to the rating of the nodes the latter incorporates resource rating based on the usefulness of the resource for a particular user. Node based rating is commonly known as Trust management or Reputation management in P2P parlance. There are some earlier works in this direction [4] [5] [6] [7] [8]. A centralized node rating approach is used in some popular networks [9] and [10]. Some researchers have proposed Global rating approaches [11] and [12], which calculate, issues and updates the rating to each node and publishes them globally. There are some local rating mechanisms based on past interactions between peers [13] and [14]. The concept of node rating is attractive because it effectively solves the problem of network security and privacy but in case of better information retrieval resource based rating is better than node rating because of the following reasons:

- The assessment of a nodes reputation is based on the interaction of that node with the members of the network involving large scale information to be processed and stored while the resource rating is simply based on the quality of the resource.

- A new peer having a low reputation in the network can have required resource but due to its low reputation it will have less chance to share that resource.

- It is comparatively simple to attach and modify ratings of a resource than modifying rating of a node because latter results from a large scale interaction while former depends on itself only.

- For some special networks such as Research and Education network, resource rating is better than node rating because in these networks peers are interested in specific information which can be served via resource rating.

Although resource rating is attractive and has some distinct advantages over node rating there has been comparatively less work done in this field. Chen and Chen [15] have proposed a network in which the requester asks other peers to rate the resources and weighs their advice according to their reliability. The rating is again based on the responses of the peers and brings them into the process making the whole process network dependent. The proposed model accounts for this by attaching ratings to the documents itself.

## 3. JXTA

Although the applications of P2P technologies are fascinating and still much more is left to be discovered but there are some challenges that are to be faced. Unfortunately, the current applications of P2P tend to use protocols that are incompatible in nature, reducing the advantage offered by gathering devices into P2P networks. Each network forms a closed community, completely isolated from the other networks and incapable of using their services.

So, the requirement of a common platform which can bind all the peers and facilitates free communication between them is pre-requisite for P2P to realize its full potential. Project JXTA is the solution to this problem provided by Sun Microsystems. JXTA is simply a set of protocol specifications [16] [17] [18]. So use of JXTA makes the production of a new P2P application very easy which is what makes it so powerful. Basically JXTA could be thought of as a programming language like C or C++. Hence, it provides a great deal of flexibility and new possibilities. The basic protocol stack of JXTA is shown in Figure 1. The JXTA is a layered application, on top which resides the Applications and at its core the peer networks ensure all the functionalities and services. A detailed layer structure of JXTA is shown in Figure 2.

JXTA provides a set of basic protocols based on which some standard libraries have been constructed. These libraries provide a good deal of application support along with the possibility of adding new features according to the requirements of a user.

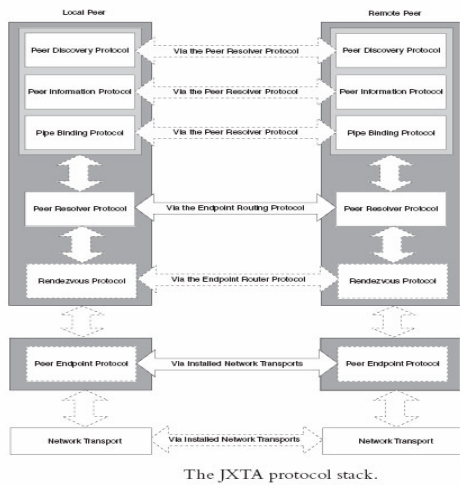 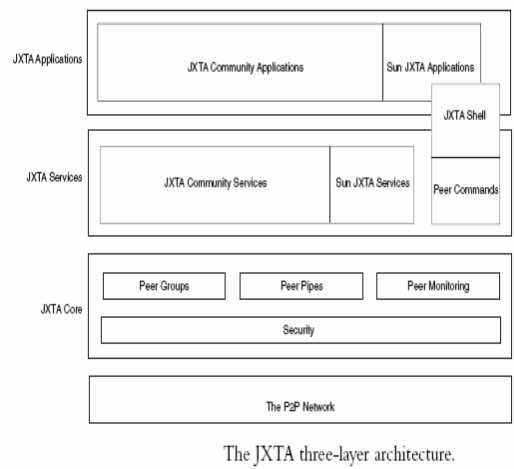

Figure 1. The JXTA protocol stack[17]       Figure 2. The detailed layer structure of JXTA[17]

## 4. MYJXTA

Although JXTA provides a SHELL command interface for the fulfilment of basic requirements like- making peer groups, joining a group, uploading data, publishing advertisements etc. yet the need of a more user-friendly and feature-rich higher level application was raised by the community using JXTA. So, myJXTA was launched as an answer which was pretty easy to use and has some extra features like chat option. The myJXTA interface which has been modified to suit the requirements according to the needs of *vuCRN* [4 2] is shown in Figure 3.

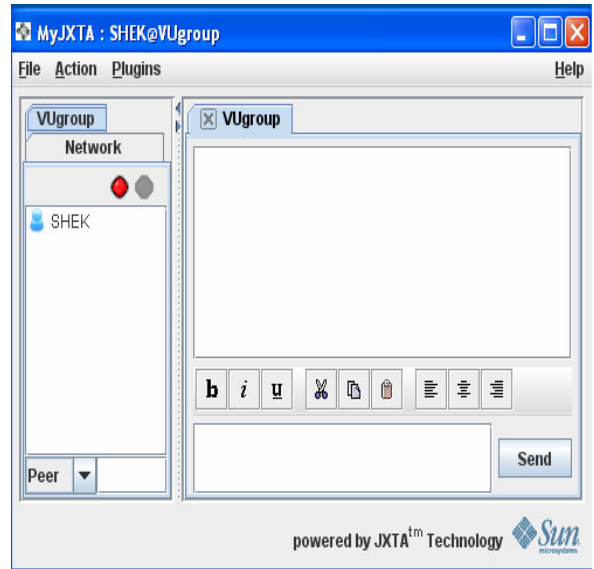

Figure 3. Interface of myJXTA for *vuCRN*

## 5. RATED-RESOURCE P2P NETWORK (R2P2P)

As mentioned already that a collaborative research network must be precise in supplying information to the user a possible model is proposed in this section to achieve the task. *vuCRN* has already achieved the task of supervised file uploading and Digital Right Management [2], [3]. The *vuCRN* network file sharing architecture is shown in Figure 4 and the reader is requested to go through the architecture of file sharing in *vuCRN* for a better understanding of the proposed model. In *vuCRN* network user authentication is required for uploading any document in the network which is controlled by LDAP server. While the document is uploaded as a PDF file on the network information regarding the copyrights is attached to it as metadata. This model further extends this idea by attaching a rating to the document which can be parsed in real-time and thus the responses are sorted accordingly. Various parameters like entity of user, description of the document are taken into consideration while deciding the final rating for the document. Without any server this task looks impossible but an innovative solution to this problem using the advertisement published by a peer is presented.

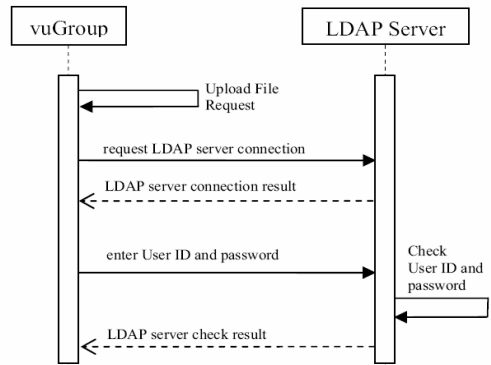
Figure 4. LDAP file sharing architecture [2]

## 5.1. Basic idea

For making a peer known to other peers in the networks JXTA provides advertisements that are issued by peers. These advertisements are in XML format and they contain all the relevant information about a peer necessary for the network functionality. These advertisements contain information about the resources shared by the peer in the network. A particular advertisement in XML is shown in Figure 5. XML is a self-descriptor language which is same as human language. Hence, the use of XML documents for advertisements makes it very easy to incorporate more features in the advertisements. Basically, a rating element <rating> is attached to the XML document which advertises the document on the network. This <rating> contains all the information regarding the relevance of the matter in form of ratings in different categories. When a peer issues a query the matching document's advertisements are parsed and displayed to the peer. In this model the <rating> element also parsed by the enquiring peer and the result of the search is sorted according to the rating attached with the document. The work of extracting information encapsulated in <rating> element and processing this information for sorting the search result is done by the querying peer.

```
JXTA>cat MyPipeAdvertisement
<?xml version='1.0'?>
<!DOCTYPE jxta:PipeAdvertisement>
<jxta:PipeAdvertisement xmlns:jxta="http://jxta.org">
    <Id>
        urn:jxta:uuid-59616261646162614E504720503250339C0C74ADD709
4CEC90EC9D4471DFED5304
    </Id>
    <Type>JxtaUnicast</Type>
</jxta:PipeAdvertisement>
```

Figure 5. A typical XML advertisement

## 5.2. Relevance parameters

It is a common experience that relevance of a particular matter is subjective and depends heavily on the user. For example, a student of B-tech level will be interested in basics of Image Processing but a Professor will be interested in state-of-art technologies and new research papers. Keeping this fact in the mind the proposed model sorts the documents according to the user. This is made possible by adding child elements in the root element <rating> that contain the information regarding the relevance according to the user. The child elements are:

- **Citations**
  It contains the information regarding the number of citations which indicates the authenticity and usefulness of the document. It contains an integer equal to the number of citations.
- **Level**
  This indicates the preferred level of the user. This contains alphabets which indicate the level of the user. Table 1 describes the mapping between the alphabet and the user entity. This sub-element guides the search by indicating the level of understanding of the user

Table 1 Mapping between USER & LEVEL

| USER | LEVEL |
|---|---|
| B-Tech Student | A |
| M-Tech Student | B |
| Research Scholar (PhD) | C |
| Professor | D |

- **Descriptor**
  This child element is an indicator of the type of the document like Tutorial, Research paper, Article about the basics etc. It contains an alphabet which can be parsed and information could be extracted. This element guides the search for a specific user-defined search. Table 2 shows the corresponding mapping.

Table 2 Mapping between doc type and Descriptor

| DOCUMENT TYPE | DESCRIPTOR |
|---|---|
| Basics | E |
| Tutorial | F |
| Research paper | G |

Based on these relevance parameters that are parsed in real time and the search query is thereby sorted in the order. The tree structure of the <rating> element is shown in the Figure 6. One important aspect of this model is that all the computation for calculating the rating is to be done by the querying peer. This ensures that the peer providing the resource is not laden by the load of computation due to other peer in the network. Hence, the peer sharing the resource with a particular peer can continue sharing or asking for resources without any extra computational load. This leads to a situation in which the person asking for some service is responsible for providing computation unlike, the traditional client/server model in which server has to bear the load. For ex- Google uses a cluster of 10,000 Unix systems for supporting the server for Google search engine.

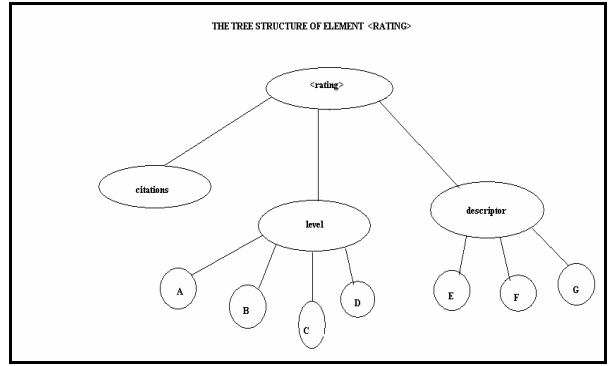

Figure 6. Tree diagram of <rating>

## 6. CONCLUSION AND FUTURE WORK

Using the proposed idea presented in this paper a flexible, user-defined, fast and robust solution can be produced to rate resources on a P2P networks. The information regarding the usability of a document is encapsulated in the <Rating> element attached to the XML advertisement of the document which can be extracted in real time. Since sorting of search result is done using the <Rating> element attached to the document by the requesting peer, the network is not overloaded. As the uploading and rating rights are reserved to few members only, the network is free from pollution and malicious contents. In the future probabilistic networks would be used for sorting the document for better sorting.

## ACKNOWLEDGEMENTS

The authors would like to thank Faculty of Health, Engineering and Science, Victoria University, Australia for providing the three months internship in 2008.

**Authors**

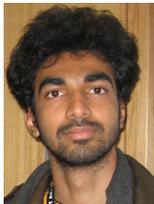

**Abhishek Sharma** is a final year student at Department of Electrical Engineering, IIT Roorkee. He is two times winner of National Level Paper Presentation Contest in India. His research interests are Computer Vision, machine learning, artificial neural networks and computer networking. He has authored and co-authored 6 articles in journals and conferences including one in Neurocomputing Elsevier and IEEE Transaction on Power Delivery. He has worked under supervision of Associate Professor Hao Shi on P2P network in 2008.

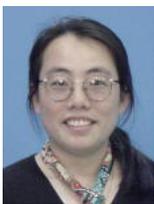

**Dr. Hao Shi** is an Associate Professor in the School of Engineering and Science at Victoria University, Australia. She completed her PhD in the area of Computer Engineering at the University of Wollongong and obtained her Bachelor of Engineering degree from Shanghai Jiao Tong University, China. She has been actively engaged in R&D and external consultancy activities. Her research interests include p2p Networks, Location-Based Services, Web Services, Computer/Robotics Vision, Visual Communications, Internet and Multimedia Technologies.